# INFLUENCE OF ANNEALING CONDITIONS ON STRUCTURE AND OPTICAL PROPERTIES OF COPPER NANOPARTICLES EMBEDDED IN SILICA MATRIX

Oleg A. Yeshchenko[1], Igor. M. Dmitruk[1], Andriy M. Dmytruk[2], Alexandr A. Alexeenko[3]

[1] *Physics Department, National Taras Shevchenko Kyiv University, 2/1 Akademik Glushkov prosp., 03127 Kyiv, Ukraine*

[2] *Center for Interdisciplinary Research, Tohoku University, Aoba-ku, Aramaki Aza Aoba, 980-8578 Sendai, Japan*

[3] *Gomel State Technical University, Gomel, Belarus*

Copper nanoparticles have been grown in silica matrix by annealing of the sol-gel prepared porous matrix impregnated with the copper nitrate. The annealing has been performed in air, successively in air and hydrogen, and in hydrogen. Cu nanoparticles in size range of 2-65 nm have been grown depending on annealing conditions. Annealing in air results in copper oxide nanoparticles ($Cu_2O$) growth as well. Transmission electron microscopy (TEM) and optical spectroscopy of the copper nanoparticles in silica matrix have been performed. The copper nanoparticles of two types are grown: spherical "mature" particles and elliptical "seed"-particles. The surface plasmon peak has been observed clearly in absorption spectra of Cu nanoparticles. Surface plasmon peak in absorption spectra of Cu nanoparticles demonstrates slight blue shift with decrease of the particle size. The half-width of the surface plasmon peak decreases appreciably at the lowering of temperature from 293 K to 77 and 4.2K that is due to strong electron-phonon interaction. The low-frequency Raman scattering data are in agreement with electron microscopy and absorption data. Photoluminescence from the copper nanoparticles has been observed. Efficiency of the luminescence increases appreciably at the decrease of particle size. The observed increase is explained, probably, by the coupling of the excited incoming and outgoing emitted photons with surface plasmon.

1. Introduction

It is well known that the optical properties, the absorption in particular, of the nanosized metals differ drastically from the properties of bulk metals. The physical origin of the light absorption by metallic nanoparticles is the excitation of coherent oscillations of the conduction band electrons by light [1]. These oscillations are known as surface plasmons that are readily observable in case of nanoparticles which have very large specific area [1,2]. However, the size dependence of the surface plasmon absorption is not as easily explained as in the case of semiconductor nanoparticles [1], where a shift of the HOMO and LUMO results in a larger band gap and a blue shift of the absorption bands with decreasing size. The extinction coefficient of small metallic particles is given in Mie's theory as the summation over all electric and magnetic multipole oscillations contributing to the absorption and scattering of the interacting electromagnetic field [1,2]. For nanoparticles much smaller than the wavelength of the absorbing light only the dipole term contributes to the absorption. In the quasi-static regime the extinction coefficient $\kappa$ for $N$ particles containing in volume $V$ is expressed by the following equation:

$$\kappa = \frac{18\pi N V \varepsilon_m^{3/2}}{\lambda} \frac{\varepsilon_2}{\left(\varepsilon_1 + 2\varepsilon_m\right)^2 + \varepsilon_2^2}, \qquad (1)$$

where $\lambda$ is wavelength of the absorbing light, $\varepsilon_m$ is the dielectric constant of the surrounding medium (assumed to be independent on frequency), $\varepsilon_1$ and $\varepsilon_2$ are the real and imaginary parts of the material dielectric constant respectively depending on the light frequency. The resonance condition for the surface plasmon absorption is roughly fulfilled when $\varepsilon_1 = -2\varepsilon_m$. This condition depends weakly [1] on $\omega$. The plasmon half-width mainly depends on $\varepsilon_2(\omega)$. According to equation (1), the plasmon absorption is size-independent within the dipole approximation. However, experimentally a size dependence of the surface plasmon absorption is observed with decrease of particle size. Since Mie's theory has found wide applicability and has generally been successful in explaining optical absorption spectra of metallic nanoparticles [1], a size dependence for the quasi-static regime is introduced in equation (1) by assuming a size-dependent material dielectric constant $\varepsilon(\omega, r)$ [1]. The dielectric constant can be written as a combination of a $\varepsilon_{IB}(\omega)$ term caused by interband transitions from electron $d$ states in valence band, and a Drude term $\varepsilon_D(\omega)$ considering the free conduction electrons only, i.e. $\varepsilon(\omega) = \varepsilon_{IB}(\omega) + \varepsilon_D(\omega)$. The latter term is given within the free electron model by the following expression [1]:

$$\varepsilon_D(\omega) = 1 - \frac{\omega_p^2}{\omega^2 + i\gamma\omega} , \qquad (2)$$

where $\omega_p$ is the bulk plasmon frequency and $\gamma$ is a phenomenological damping constant and equals the surface plasmon half-width for the case of a perfect free electron gas in the limit of $\gamma \ll \omega$. Experimentally, both a blue shift and a red shift of the plasmon peak with decreasing particle size have been reported earlier [1]. A variety of theoretical approaches considering different small particle effects (e.g., size-dependent changes of the electron band structure causing the blue shift, "spill out" of conduction electrons causing the red shift, etc.) reach contradicting conclusions for the dependence of the plasmon peak position on the nanoparticle size.

Nanosized copper together with other noble metals as gold and silver are the most studied metallic nanoparticles as the surface plasmon resonances are clearly featured in the optical spectra, and are located in visible region. Among the most studied Cu nanoparticles are ones embedded in an insulating matrix [1,3]: optical properties of copper nanoparticles have been widely investigated in a wide range of host matrices [4–7], including glasses [8-11]. However, data on the optical properties of copper species in pure silica glass are seldom because of the difficulties encountered in the preparation of this kind of materials. On the other hand, the sol-gel method applied to silica is well known to enable the preparation of compounds that are not accessible by standard ways of synthesis. Moreover, products of very high optical quality and homogeneity can be obtained. This is essential for optical studies and applications.

In this paper we report the synthesis and characterization of the copper nanoparticles in silica matrix by methods of transmission electron microscopy (TEM) and optical spectroscopy (absorption, Raman scattering and photoluminescence). The size and shape distributions of Cu nanoparticles are determined from the data of TEM and optical spectroscopy.

## 2. Nanoparticle growth and sample preparation for optical measurements. Experimental procedures

Technology of the growth of Cu nanoparticles in $SiO_2$ matrix is similar to described in Ref. [9]. Porous silica matrices (obtained by the transformation to monolithic glasses) were produced by the conventional sol–gel technique based on hydrolysis of tetraethoxysilane (TEOS). We have modified the sequence introducing a doping followed by the chemical transformation of the dopants in air or controlled gaseous medium. A precursor sol was prepared by mixing of TEOS, water and alcohol, with the acid catalysts $HNO_3$ or HCl. Silica powder with particle size about 5-15 nm (the specific surface area is $380 \pm 30$ m$^2$/g) was added into the sol followed by ultrasonication procedure. The key role of this procedure was to prevent a large volume contraction during drying. The next gelation step resulted in formation of gels of desired shape. Gels were dried at room temperature under humidity control, and porous materials (xerogels) were obtained at this step. Their annealing in air at the temperature of 600 $^o$C during 1 h allowed us to regulate porosity of $SiO_2$ matrices. This heating process resulted also in the noticeable density and specific surface area variations. Copper doping was performed by impregnation of xerogels in $Cu(NO_3)_2$ alcohol solution during 24 h. Then, the impregnated samples have been dried in air at 40 $^{o}$C during 24 h. The further processing of the Cu-doped xerogels was done by the three pathways. (1) An annealing in air with gradual increase of the temperature from 20 $^{o}$C to 1200 $^{o}$C (annealing time at 1200 $^{o}$C: 5 min). (2) The initial annealing in air with gradual increase of the temperature from 20 $^{o}$C to 1200 $^{o}$C (annealing time at 1200 $^{o}$C: 5 min), then annealing in the atmosphere of molecular hydrogen at the temperature of 800 $^{o}$C during 1 h. (3) An annealing in $H_2$ with gradual increase of the temperature from 20 $^{o}$C to 1200 $^{o}$C (annealing time at 1200 $^{o}$C: 5 min). As we show below, an annealing in air leads in general to creation of the copper oxide ($Cu_2O$) nanoparticles although the small Cu nanoparticles with low concentration in the sample grow as well. An annealing in hydrogen results in reduction of Cu(I,II) to Cu(0) that is aggregated in the form of Cu nanoparticles. Glass samples fabricated were polished up to thickness about 1 mm for optical measurements. The sets of samples were prepared. The samples obtained by annealing in air have low optical density, and are non-colored or slightly yellow (which are labeled hereafter as samples A); ones obtained at successive annealing in air and hydrogen have higher optical density, and have orange-red or red color (samples AH); ones obtained at annealing in hydrogen have the highest optical density, and have red color (samples H). After removals of a layer with thickness 0.5-1 mm from samples AH, the samples become light-pink colored. It indicates that at this reduced treatment the transformation from copper oxide to reduced copper occurs non-uniformly on depth of the sample.

The optical spectra of the samples placed in air, liquid nitrogen or liquid helium have been measured. A tungsten-halogen incandescent lamp has been used as a light source for the absorption measurements. An $Ar^+$ laser with wavelength 514.5 nm has been used in Raman experiments, and $Ar^{++}$ laser ($\lambda$ = 351.1 nm) has been used for the excitation of the photoluminescence. The single spectrometer MDR-3 has been used for the measurement of absorption and photoluminescence spectra, and the double spectrometer DFS-24 – for Raman experiments.

### 3. Transmission electron microscopy of Cu nanoparticles

Transmission electron microscopy (JEOL, JEM-2000EX) of Cu nanoparticles in $SiO_2$ was performed to determine their size and shape. Some images of the nanoparticles synthesized at the different technology conditions are presented in fig.1. The images prove the creation of the Cu nanoparticles of different sizes and shapes in the matrix in dependence of the conditions of synthesis. It should be noted the Cu nanoparticles of two types are created. First type is the large spherical nanoparticles with Gaussian-like size distribution, and the second one is the small elliptical particles with non-Gaussian size distribution. Most probably, that the large spherical particles are "mature" ones (below in the text – mature particle or MP) at the respective conditions of growth, and the small ones are some "immature seeds" (below in the text – seed particle, seed or SP) from which the large "mature" particles grow. The size of mature nanoparticles depends on the conditions of samples annealing. The copper nanoparticles of small size (diameter $d$ is in the range of 2 - 15 nm) grow in the samples annealed in the air (samples A). The medium-sized nanoparticles ($d$ = 15 - 40 nm) grow in the samples initially annealed in the air and then in the atmosphere of molecular hydrogen (samples AH). The largest particles ($d$ = 40 - 65 nm) grow in the samples annealed in the $H_2$ (samples H).

The fig.2 presents the size and shape distribution of the copper nanoparticles. It is seen that the mature particles are characterized by Gaussian-like distribution with clear-featured maximum. The seed particles have distribution monotonously decreasing with increase of particle size. One has to be noted that the seed particles are characterized by larger dispersion than mature particles. Indeed, for sample A1 annealed in air we have: $d = (6.1 \pm 0.2)$ nm ($\varepsilon_d = \Delta d / \langle d \rangle = 3.3\%$) for mature particles, and $d = (2.4 \pm 0.2)$ nm ($\varepsilon_d = 8.3\%$) for seeds; and for sample AH1 annealed successively in the air and hydrogen we have: $d = (33.9 \pm 0.8)$ nm ($\varepsilon_d = 2.4\%$) for mature particles, and $d = (10.3 \pm 1.4)$ nm ($\varepsilon_d = 13.6\%$) for seeds. Let us note that small seeds, e.g. in sample A1, are mostly spherical, and with increase of size the shape of seed-particles becomes more and more elliptical. We discuss the mechanism of the creation of mature particles from the seeds in the next section. That is seen on the inset of fig.2(c), where the dependence of the aspect ratio of the elliptical seed particles on their mean diameter is shown. Such increase of the aspect ratio with increase of size can be explained in such way. As it well known, the surface tension forces attempt to minimize the free surface of the particle making their shape spherical. The surface tension is higher for small particles (dependence $1/d$), and, therefore, the surface tension forces form the small particles into spherical shape more efficiently than the large ones.

The TEM images give an opportunity to understand the evolution of the seeds. Initially, the very small seeds of copper appear (fig. 1(b)). Then the seeds grow, their shape becomes elliptical (fig. 1(f)). The next-to-last step of mature nanoparticle formation is the creation of some dimer or trimer, constituting of two or respectively three close neighbouring large seeds. Hereafter, these large seeds merge and form the mature spherical copper nanoparticle (fig. 1(a,e,g,h)). Let us note that the seeds in fig. 1(f) locate in some capsule which contrast is lower than contrast of copper particles but is higher than contrast of the matrix. Most probably, this capsule is the $Cu_2O$ particle whose creation in the silica matrix at the annealing in the air has been reported, e.g. in Refs. [8,12,13]. Figs. 1(c,d) are very interesting, as one can see that in some samples annealing successively in the air and hydrogen the spherical copper nanoparticles with shell-like structure grow. Most probably, such shell particles have Cu shell and $Cu_2O$ core. The mean outer diameter of shell Cu nanoparticles is 18.4 nm, and the mean inner diameter is 6.2 nm that is 33.7 %, i.e. the core size is $1/3$ of the shell size.

4. Absorption spectroscopy of Cu nanoparticles

Absorption spectra of the samples containing the copper nanoparticles in $SiO_2$ matrix have been measured at the temperatures of 293K, 77 K, and 4.2 K. Absorption spectra (fig. 3(a)) demonstrate clear surface plasmon (SP) peaks located in the region 560.8 nm (2.210 eV) – 581.1 nm (2.133 eV) for different samples. The SP peak spectral positions are in good agreement with the values predicted by Mie theory for the Cu nanoparticles of the corresponding sizes embedded in silica matrix. Taking the diameters of the nanoparticles from TEM data we have obtained the dependence of the half-width of plasmon peak on the nanoparticle diameter. The respective dependence is shown in fig. 3(b). It is seen that the dependence of the half-width on size is described very good by the well known $1/r$ law [1] which gives the size-dependence of damping constant of plasmon that is due to limitation of mean free path of electrons caused by scattering of the electrons by the surface of the nanoparticles:

$$\gamma = \gamma_\infty + A \frac{v_\infty}{r}, \qquad (3)$$

where $\gamma$ is half-width of plasmon absorption peak, $\gamma_\infty = 1/\tau_\infty$ is the size-independent damping constant which is related to the lifetimes of electron-electron, electron-phonon, and electron-defect scattering, $v_\infty$ is the Fermi velocity in bulk material ($v_\infty = 1.57 \times 10^8$ cm/sec in bulk copper), $r$ is the radius of nanoparticle, and $A$ is a theory-dependent parameter that includes details of the scattering process (e.g., isotropic or diffuse scattering [1,14,15]. Another quantum mechanical model considers of adsorbed molecules on the nanoparticle surface [16]. It is suggested that the surface plasmon energy is transferred into excitation modes of the surface metal-adsorbate complex (chemical interface damping). Many more theories exist [1,17-19], which all find a $1/r$ dependence, reflecting the importance of the ratio between the surface area and the volume. However, the parameter $A$ depends on the theory. We have performed the fitting of our experimental dependence $\gamma(d)$ obtained at the temperature of 293 K. The fitting gives the following values of parameters from expression (3): $\gamma_{\infty,293} = 0.087$ eV and $A_{293} = 0.107$.

The similar dependence $\gamma(d)$ has been determined from the absorption spectra measured also at the temperature of 77 K (fig. 3(b)). It can be seen that the half-width of plasmon peak decreases strongly at decrease of the temperature (see fig. 3(a)). The plasmon peak is narrowing strongly at the decrease of temperature from 293 K to 77 K. However, at the further decrease of the temperature from 77 K to 4.2 K the half-width of the plasmon peak does not change. Such broadening of the plasmon peak at the increase of the temperature is due to the increase of the frequency of electron-phonon scattering acts with increase of temperature. The inset of fig. 3(b) shows the dependence of the ratio of the plasmon half-widths at the temperatures of 293 K and 77 K. It is seen that this ratio increase appreciably at the increase of the nanoparticle size. Apparently, it reflects the fact of the strengthening of the electron-phonon interaction at increase of the particle size. The fitting of our experimental dependence $\gamma(d)$ obtained at the temperature of 77 K gives the following values of parameters from expression (3): $\gamma_{\infty,77} = 0.036$ eV and $A_{77} = 0.103$. One can see that the parameter $A$ is the same for the temperatures of 293 K and 77 K indicating the facts that the scattering of electrons on the boundaries of the nanoparticle are non-dependent on the temperature. The value of the size-independent damping constant $\gamma_\infty$ lowers in 2.4 times at the decrease of the temperature from 293 K to 77 K reflecting the facts of the existence of strong electron-phonon interaction in studied Cu nanoparticles embedded in silica matrix.

A small blue shift of plasmon peak with the decrease of the Cu nanoparticle size is observed in the absorption spectra (see figs. 4 and 5). The observation of the blue shift in our samples is an evidence of the small role of the "spill out" effect. Most probably, the observed blue shift is the manifestation of the change of the electron band structure with the change of the Cu nanoparticle size. The fig. 5 shows the plasmon peak energy dependence on the nanoparticle diameter for the temperatures of 293 K and 77 K. One can see that unlike to the strong temperature dependence of plasmon peak half-width, within errors of absorption peak position determination no dependence of plasmon peak position on the temperature is observed. This fact indicates that the structure of Cu nanoparticles does not change at the change of temperature from low one up to 293 K.

It is seen from the fig. 4(b) that the plasmon band has the two-component structure. As we note in section 3, the TEM shows that the creation of the Cu nanoparticles of two types takes place in our samples. First ones have been called as mature particles (MP), and the second ones – seed particles (SP). Therefore, one can assume that the narrow low-energy absorption band is caused by the absorption in mature Cu nanoparticles, and the high-energy wide one is caused by the absorption in seed Cu nanoparticles. One can see from the figs. 4(b) and 5 that bands are blue shifted with decrease of the particle size, but the change of the SP band position with the particle size occurs faster than the one for the MPs. The spectral split of the absorption bands of mature and seed particles decreases with the increase of particle size, and disappears for largest particles with the diameters $\geq 60$ nm. A red shift with the increase of size of the SP's high-energy band is similar to the red shift of MP's low-energy one. The fact of the stronger dependence of SP's band position on the size can be explained in such a way. With increase of the size the distance between the seeds becomes equal or smaller comparing to their size (fig. 1(f)). Respectively, the interaction between the particles-seeds becomes quite strong, that results in the lowering of their energy. At the further increase of size of seeds they merge forming the mature particle. Meanwhile, the distance between the mature particles remains a lot larger than their size (see fig.1 (g)). Consequently, the interparticle interaction for MPs does not play such important role as for SPs. So, for the SPs both effects of the size dependent red shift and the energy lowering caused by interparticle interaction with increasing size take place, for MPs only an effect of the size dependent red shift with increasing size takes place. The fact that the half-width of SPs is larger than one for MPs is due to following causes. Firstly, the SPs are characterized by more broad size and aspect ratio distributions comparing to the distribution of MPs as we note in section 3. Secondly, the seed particles are characterized by larger values of half-width than ones of mature particles accordingly to $1/r$ law expressed by (3). The similar absorption spectra with two-component structure of surface plasmon resonance of the copper nanoparticles in colloidal solutions have been reported earlier [20]. Authors of Ref. [20] report the tuning of the optical (absorption) properties of shaped Cu nanoparticles as nanodisks, elongated nanoparticles, and cubes with their shapes.

As we note in a section 3, on the TEM images of the samples annealed in air only and samples annealed successively in air and hydrogen the seeds particles are located in some capsules contrast of which is higher than the contrast of surrounding matrix, but lower than contrast of Cu nanoparticles. As we note in the previous section, one can assume that these capsules are the particles of copper oxide $Cu_2O$. The $Cu_2O$ particles absorb the light causing an appearing in the absorption spectrum of two bands marked as $O_1$ and $O_2$ (O – oxide). The positions of the maxima of these bands are 450 nm ($O_1$) and 475 nm ($O_2$) that is in good agreement with positions of the absorption bands of $Cu_2O$ particles in silica matrix: 455 nm and 478 nm [8]. The 475 nm band can be attributed to the blue-shifted $n=1$ exciton transition (612.5 nm at 77 K in bulk $Cu_2O$) of the yellow series of $Cu_2O$ particles. Correspondingly, the band at 450 nm can be attributed to the blue shifted $n=1$ exciton transition (580 nm at 77 K in bulk $Cu_2O$) of the green series. One can see from the fig. 4(a) that the relative intensity of $Cu_2O$ absorption bands ($O_1$ and $O_2$) decreases and the relative intensity of Cu bands increase with the increase of Cu particle size. This dependence is the result of the decrease of quantity of the $Cu_2O$ with increase of the level of Cu reduction into the metallic state. Let us note that the $O_1$ and $O_2$ bands of copper oxide are more intensive than the intensity of the Cu surface plasmon P band in the samples annealed in air only. Relative intensity of $O_1$ and $O_2$ bands decreases appreciably in the samples annealed successively in air and hydrogen. Lastly, the $O_1$-$O_2$ bands are absent in the samples annealed in hydrogen only.

## 5. Raman spectroscopy of Cu nanoparticles

Low-frequency Raman spectroscopy is the tool that is used frequently for the determination of the size of nanoparticle. Many reports on the observation of the low-frequency Raman scattering of nanoparticles embedded into transparent matrices exist, e.g. for semiconductor [21,22] and metallic [10,23,24] particles. These Raman peaks were attributed to acoustic-phonon modes confined in homogeneous particles. The energies of these vibration eigenmodes depend on the elastic properties of the material (longitudinal and transverse sound velocities) and are also a function of the particle diameter. In a simplified model [25] that considers vibrations of a free homogeneous elastic particle only two types of acoustic modes are expected to be Raman active. Those are the symmetric spheroidal mode (the so called "breathing mode") labeled $l=0$ and asymmetric quadrupolar one $l=2$. The classification of these modes is performed following the symmetry group of a sphere, with the indices $l$ and $m$ equivalent to the spherical harmonics $Y_{lm}$. The frequencies of spheroidal "breathing" mode $\omega_0$ and quadrupolar one $\omega_2$ are expressed [25] as following:

$$\omega_0 = A_0 \frac{v_L}{dc}, \tag{4}$$

$$\omega_2 = A_2 \frac{v_T}{dc}, \tag{5}$$

where $\omega_0$ and $\omega_2$ are the frequencies of the modes with $l=0$ and $l=2$, $v_L$ and $v_T$ is the longitudinal and transversal sound velocities, $A_0$ and $A_2$ are proportionality coefficients depending on the angular momentum $l$, the harmonic number $n$, and the ratio between the longitudinal and transverse sound velocities, $c$ is the velocity of light in vacuum, and $d$ is the diameter of the vibrating sphere. From the frequency position of the low-frequency Raman peaks, one can derive the mean particle size. The values of $A_0$ and $A_2$ coefficients have been calculated in Ref. [25] as a function of $v_L/v_T$. Using the results of Ref. [25] and the sound velocities for bulk copper $v_L$ = 4828 m/sec and $v_T = 2648$ m/sec, one can be obtained the value $A_0 = 0.83$.

We have measured the low-frequency Raman spectra of the Cu nanoparticles in SiO$_2$ matrix. Despite of the low intensities of the observed peaks, the stokes Raman spectra of studied samples have been measured. Some of these spectra are presented in fig. 6. The observed peaks have the following frequencies: 16.0 cm$^{-1}$ for the sample containing Cu nanoparticles with mean diameter of 6.1 nm, 15.3 cm$^{-1}$ (5.1 nm), 4.5 cm$^{-1}$ (34.5 nm), 3.9 cm$^{-1}$ (43.7 nm), and 3.2 cm$^{-1}$ (47.6 nm). Assuming the observed peaks originated from the spheroidal "breathing" $l=0$ mode of Cu nanoparticles, and proceeding from the expression (4), we have estimated the diameters of the particles. The obtained respective values are the following: 8.3 nm, 8.8 nm, 30 nm, 34 nm, and 41.7 nm. The diameters obtained from the Raman spectra are in satisfactory agreement with the values determined from TEM and absorption.

## 6. Photoluminescence of Cu nanoparticles

The photoluminescence (PL) spectra of Cu nanoparticles in silica matrix have been studied at the temperatures of 293 K and 77 K. Some of obtained spectra are presented in fig. 7(b). Additionally, we have studied the PL spectra of both non-annealed and annealed undoped $SiO_2$. The respective spectra are shown in fig.7(a). It can be seen from the fig. 7(a) that the photoluminescence of non-annealed $SiO_2$ is located mainly in violet-blue-turquoise region (380-480 nm) with the maximum at 430 nm. The respective band is marked as M (M – matrix) in the fig. 7. Through the annealing at the conditions used for the Cu nanoparticles formation the low-energy bands appear in the PL spectra of undoped $SiO_2$ in red region. We have marked these bands as $MA_1$ and $MA_2$ (MA – matrix annealed) respectively. Their maxima are located at 600 nm and 670 nm. At the lowering of the temperature from the 293 K to the 77 K the intensity of these bands decreases drastically, and the single M band become the dominant in the spectra. The behaviour of the PL spectra described above is evidence that the $MA_1$ and $MA_2$ bands appearing in the spectra at the annealing are caused by some defect centers where the electronic excitations can be localized. As the intensity of these bands decreases sufficiently at the decrease of temperature, one can draw a conclusion that the emission from these defect centers occurs only at their thermoactivation, i.e. some thermoactivated photoluminescence takes place.

The PL spectra of the annealed $SiO_2$ containing Cu nanoparticles are presented in fig. 7(b). One can see that the spectra contain the bands of the matrix (M, $MA_1$ and $MA_2$ bands), as well as new band assigned to the emission from the Cu nanoparticles. We mark this band as Cu. The maximum of PL band of Cu nanoparticles is 560.1 nm (2.213 eV). The Cu band overlaps highly with the MA$_1$ band of annealed SiO$_2$ which have the defect origin as we have mentioned in the preceding paragraph. At the lowering of the temperature to 77 K the intensity of the defect matrix MA$_1$ and MA$_2$ bands decreases drastically, and as result the Cu band can be extracted from the total spectrum. It can be seen best in the PL spectrum of the sample with 5 nm Cu nanoparticles in fig. 7(a)). The PL band marked as Cu can not be attributed to the luminescence of Cu$_2$O nanoparticles as the positions of Cu$_2$O nanoparticle absorption are 450 nm and 475 nm that is too higher on energy than Cu band position (560.1 nm). So large Stokes shift of about 100 nm would be expected for very small (subnanometer) clusters and molecules only and not to nanoparticles in our samples with diameters of about 10 nm.

Photoluminescence from copper and gold metals was first observed by Mooradian [26], and has been used extensively in characterizing carrier relaxation and the band structure of metals [26–29]. They found that the emission spectra excited by a 2W cw laser beam at 488 nm is totally unpolarized and did not depend on the polarization of the incident laser beam. The emission peak was centered near the interband absorption edge of the metals and was attributed to direct radiative recombination of the conduction band electrons with holes in the d-band that have been scattered to momentum states less than the Fermi momentum. The quantum efficiency was found to be on the order of 10$^{-10}$. However, many of the essential details of this mechanism, such as the excited electron and hole population distributions and the specific regions in the Brillouin zone where the recombination takes place have not been discussed. Later, Boyd et al. [30] determined the relation between the PL peaks and the interband recombination at selected symmetry points in the Brillouin zone. They also studied the effect of roughness on the surface local fields and on the PL spectra. A theoretical model to explain the radiative recombination in noble metals was developed by Apell et al. [29]. The observation of the photoluminescence of the noble metal nanoparticles and nanoclusters has been reported, e.g. in Refs. [31–34].

Mooradian at al. [26] reported for the bulk copper the maximum of PL band at 2.11 eV that is somewhat lower on energy than the maximum of the Cu band observed in the PL spectra of our samples 560 nm (2.21 eV). The proximity of these values proves our assumption that Cu band has to be attributed to the emission from the copper nanoparticles. The 0.1 eV shift of Cu-band to higher energies comparing to the position of bulk copper PL band we discuss in the next paragraph. It has been noted in Refs. [29-31] that excitation of the metal nanoparticles leads to excitation of the surface plasmons (i.e. coherent electronic motion) as well as the d-electrons. Relaxation of these electronic motions followed by the recombination of the sp-electrons with holes in the d-band leads to the observed emission. Thus, the photoluminescence of Cu nanoparticles in silica matrix observed in our experiments is due to the recombination of sp-electrons in conduction band with the holes in valence band.

One can see from the fig. 7(b) that the relative intensity of Cu band increases at the decrease of Cu nanoparticle size, i.e. the photoluminescence efficiency increases at the decrease of the size. In other words, the small metal nanoparticles "lighten" better than the large ones. According to the theoretical studies of the photoinduced luminescence [30] and the surface enhanced second-harmonic generation [35] from rough surfaces of noble metals, the incoming and outgoing fields are shown to be enhanced via coupling to the local plasmons. As well, the surface plasmon enhancement of photoluminescence from semiconductor CdS nanoparticles [36] and GaAs/AlGaAs quantum wells [37] have been observed. Such resonant interaction of surface plasmon with the photons leading to above mentioned coupling has to shift the energy of emitted photons to the energy of surface plasmon resonance. So, the coupling of the incoming and outgoing photons to the plasmon is a cause of the shift of Cu nanoparticles PL band on 0.1 eV to higher energies from the position of maximum of PL band of bulk copper. Therefore, one can assume that the increase of the efficiency of the photoluminescence of Cu nanoparticles, which is observed in our samples, is due to the coupling of the exciting (incoming) photon and emitted (outgoing) photon to the surface plasmon. Since the photons couple to the surface plasmons, an effect of efficiency increase has to be stronger in the particles where the relative contribution of the surface effects versus volume ones is higher, i.e. in the small particles. Thus, the Cu band in the observed PL spectrum is caused by the radiative recombination of the conduction band electrons with valence band holes, and the increase of the photoluminescence efficiency with the decrease of Cu nanoparticle size is due to the great enhancement of the incoming exciting light and the outgoing emitted light via the coupling of the light to the surface plasmon.

**Acknowledgments**

We thank Prof. Nicolas Dmytruk and Dr. Anatoliy Pinchuk for the fruitful and helpful discussions.

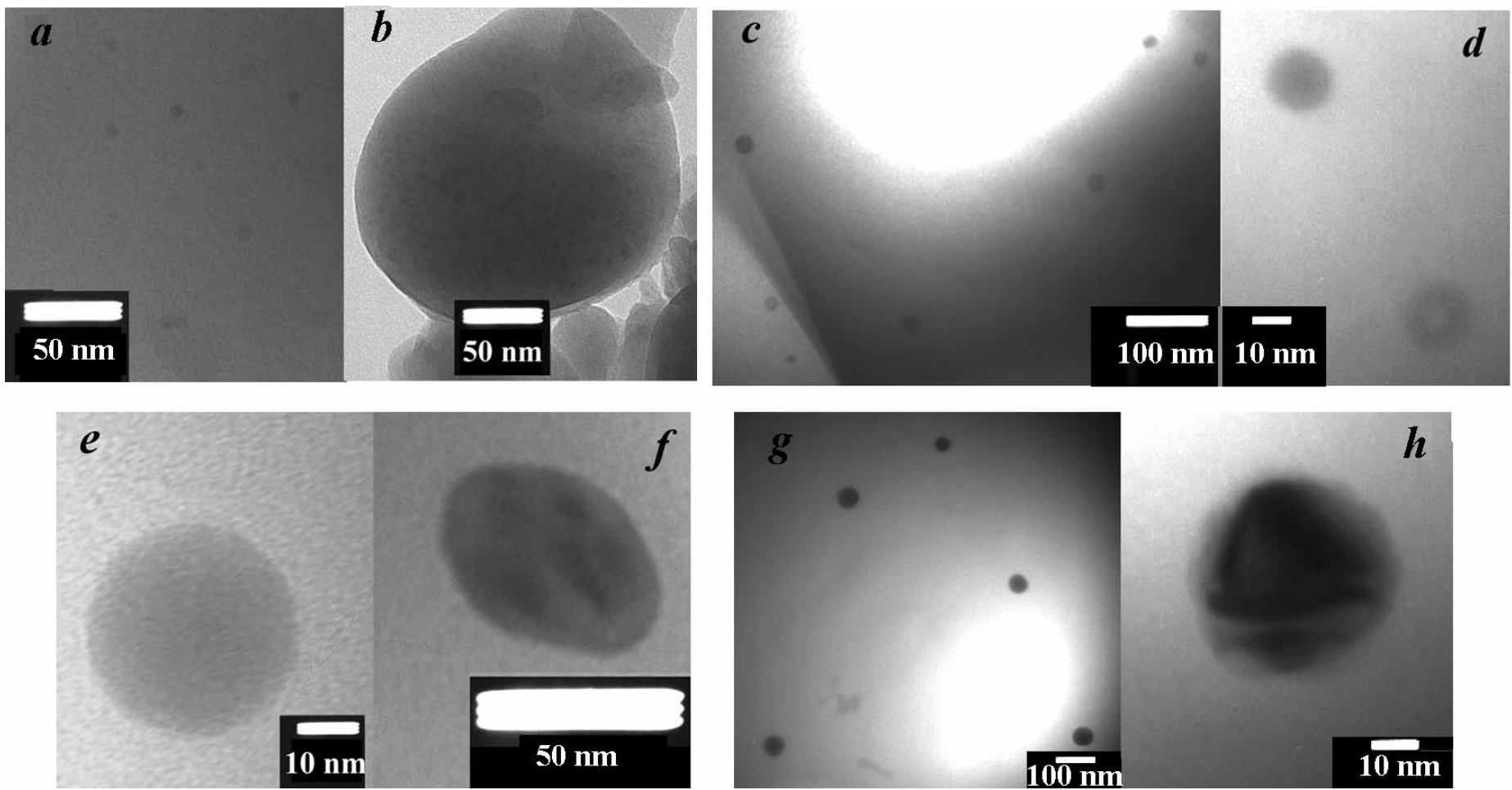

**Fig.1** TEM images of Cu nanoparticles in silica matrix. *a, b* – images of A1 sample annealed in the air; *c, d* and *e, f* – images of the samples AH2 and AH1 respectively initially annealed in the air, and after annealed in the atmosphere of hydrogen; *g, h* – images of the sample H1 annealed in the atmosphere of hydrogen.

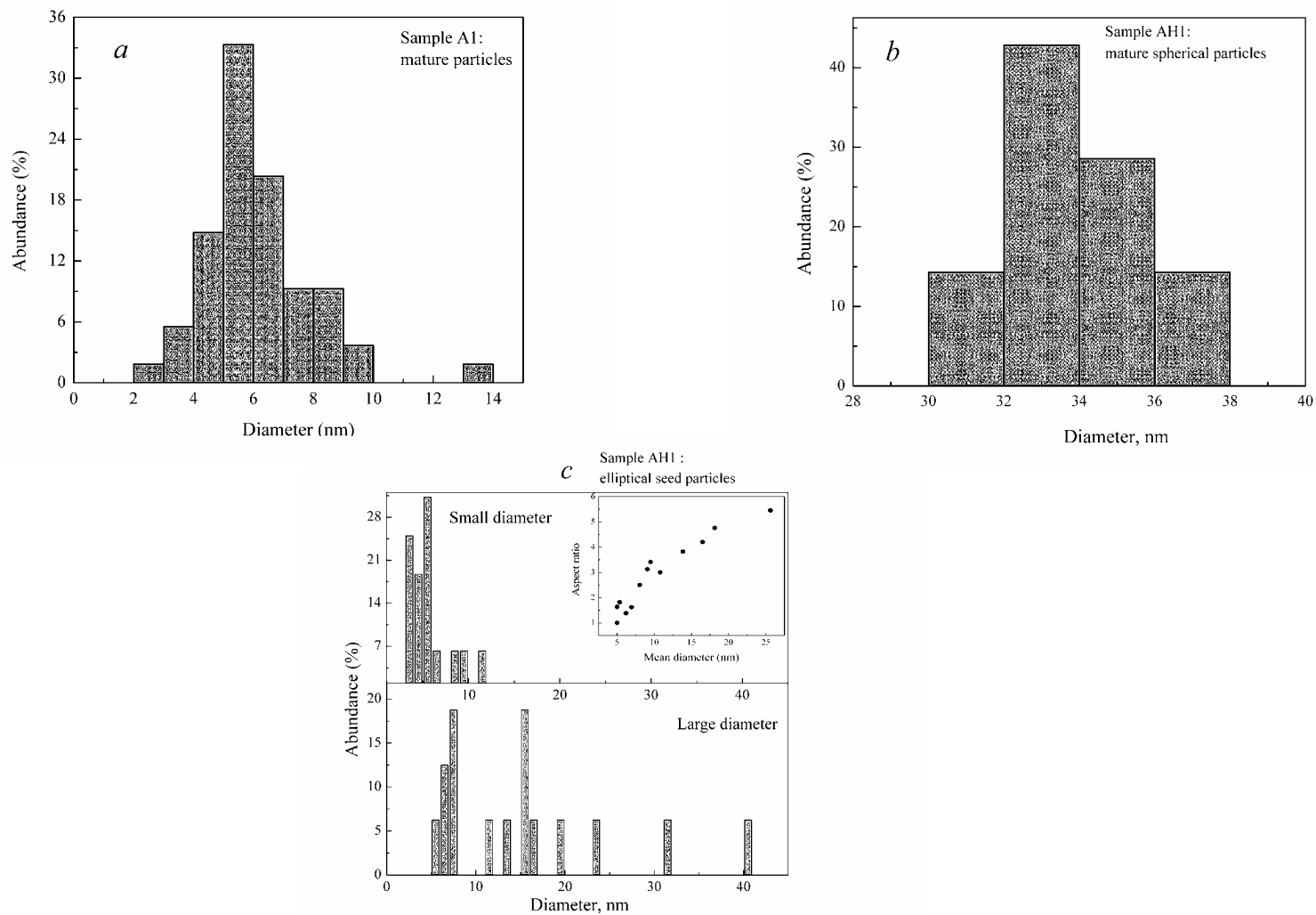

**Fig.2** Size distribution of mature (a) Cu nanoparticles in sample A1; size distribution of mature spherical (b) and seed elliptical (c) Cu nanoparticles in sample AH1. Inset in part (c) shows the dependence of the aspect ratio of seed elliptical particles in sample AH1 on their mean diameter.

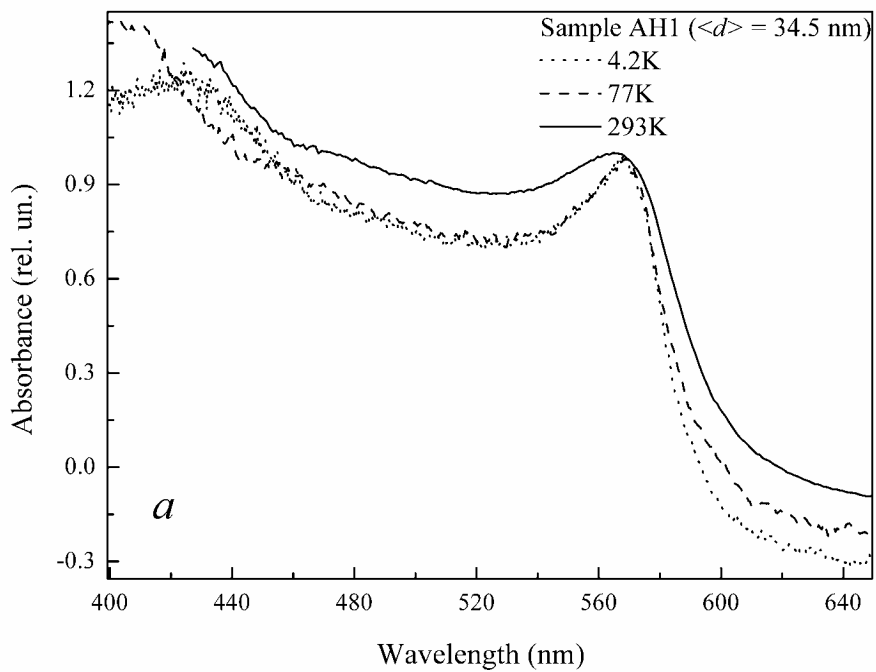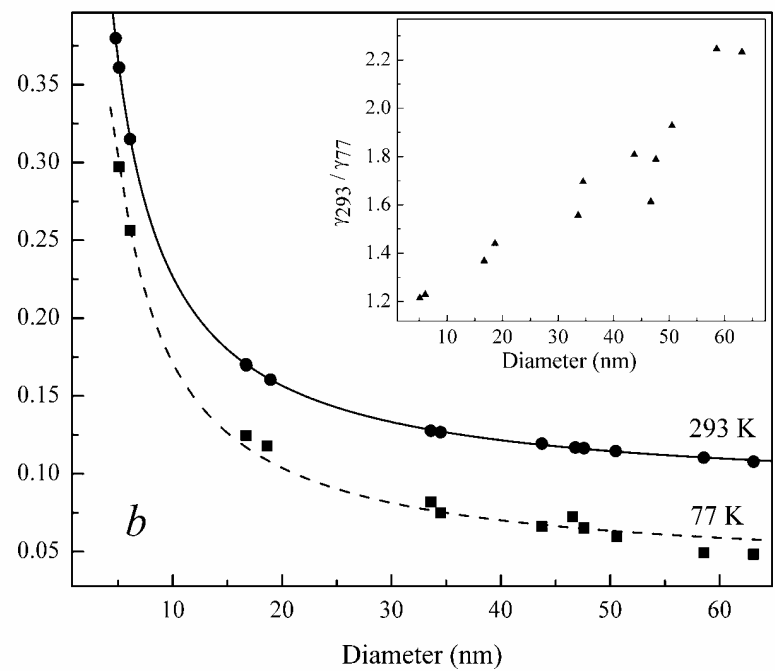

**Fig.3** Absorption spectra of Cu nanoparticles in $SiO_2$ matrix at temperatures of 293 K, 77 K, and 4.2 K (a). The dependence of the half-width of plasmon absorption peak at the temperatures of 293 K and 77 K on the diameter of nanoparticles (b); the solid line is the fitting of experimental points obtained at temperature of 293 K by expression (3), the dashed line is the fitting of experimental points obtained at temperature of 77 K by expression (3). Inset in part (b) shows the dependence of the ratio of plasmon peak half-widths at the temperatures of 293 K and 77 K on the diameter of particles.

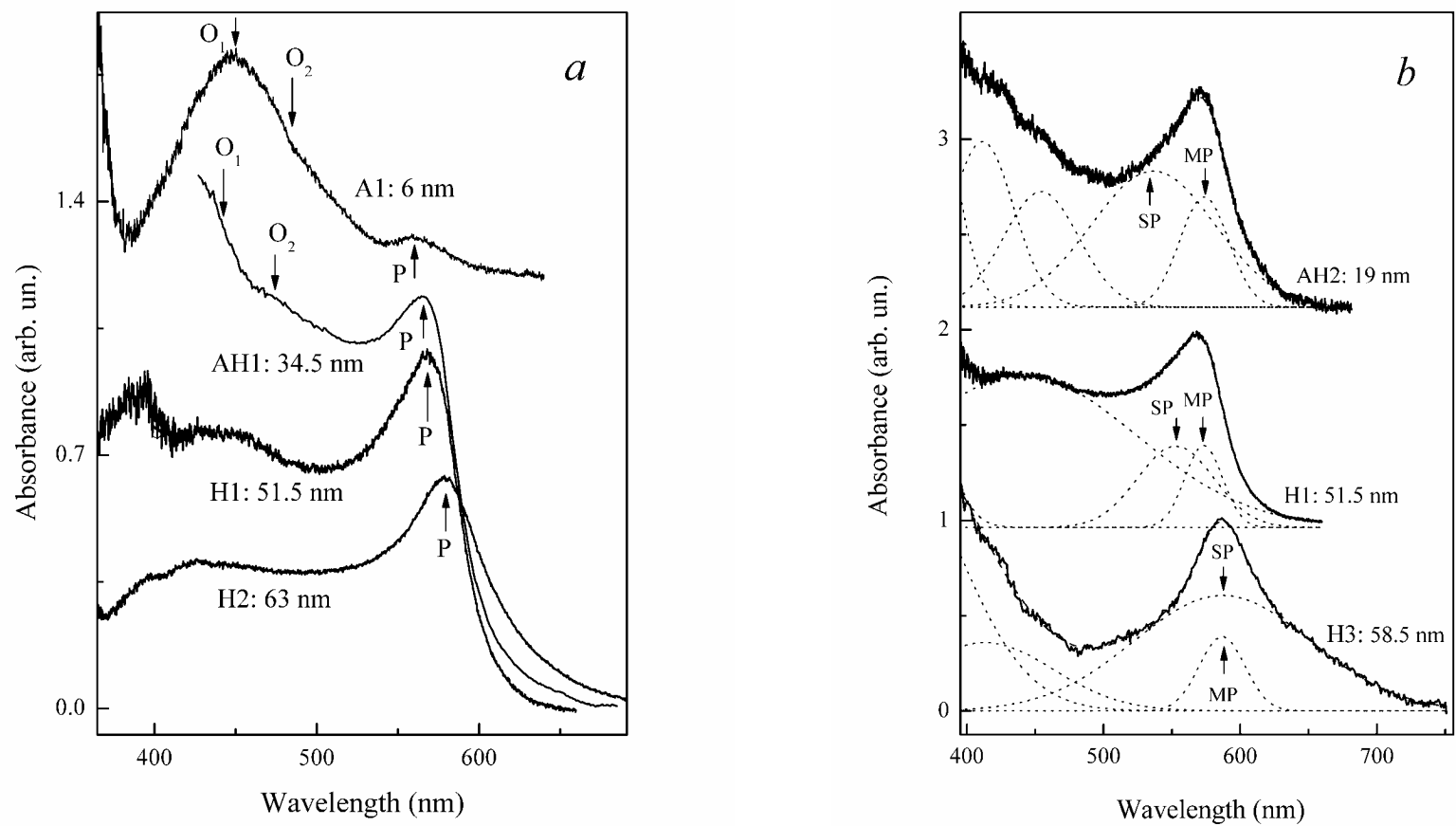

**Fig.4** (a) – Absorption spectra of Cu nanoparticles in silica matrix at the temperature 293 K demonstrating the blue shift of the surface plasmon peak with decrease of the particle size. P marks the absorption band of the surface plasmon peak in Cu nanoparticles, $O_1$ and $O_2$ are the absorption bands of $Cu_2O$. (b) – Absorption spectra of Cu nanoparticles in $SiO_2$ matrix at the temperature 293 K demonstrating the spectral split of the plasmon peaks of mature (MP) and seed (SP) particles.

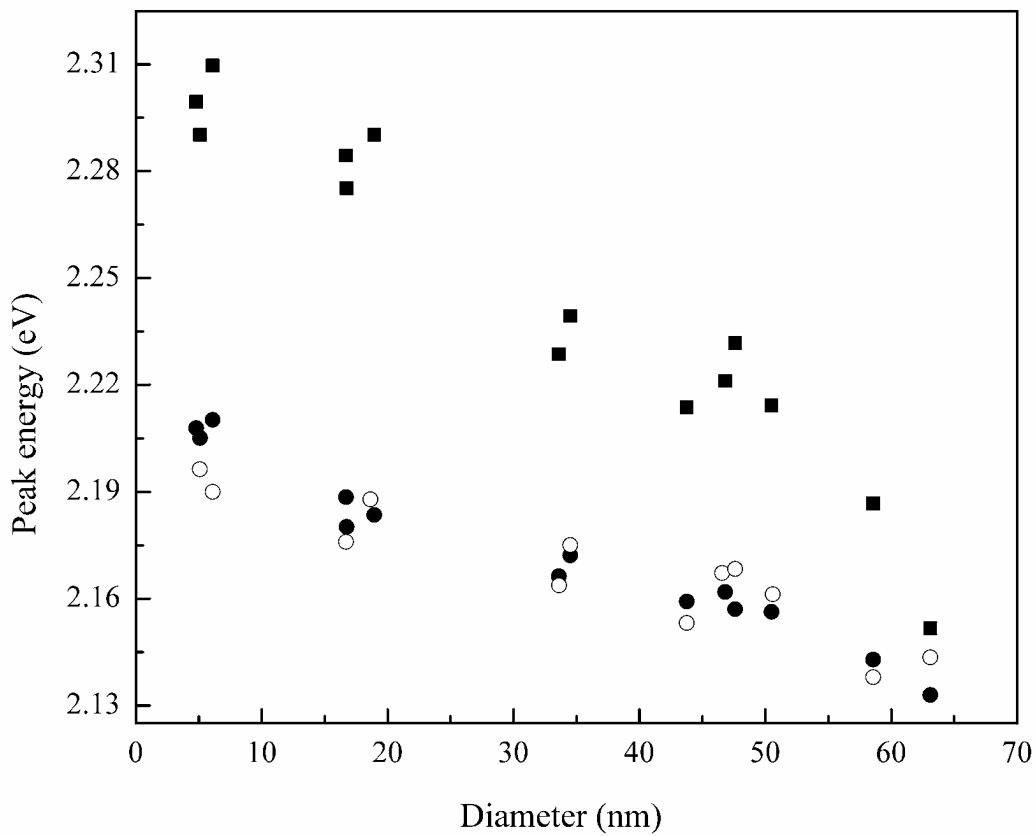

**Fig.5** Dependence of the energy of plasmon peak of mature Cu nanoparticles (MPs) at the temperature 293 K (filled circles), 77 K (open circles), and energy of plasmon peak of seed Cu nanoparticles (SPs) (filled squares) at the temperature of 293 K.

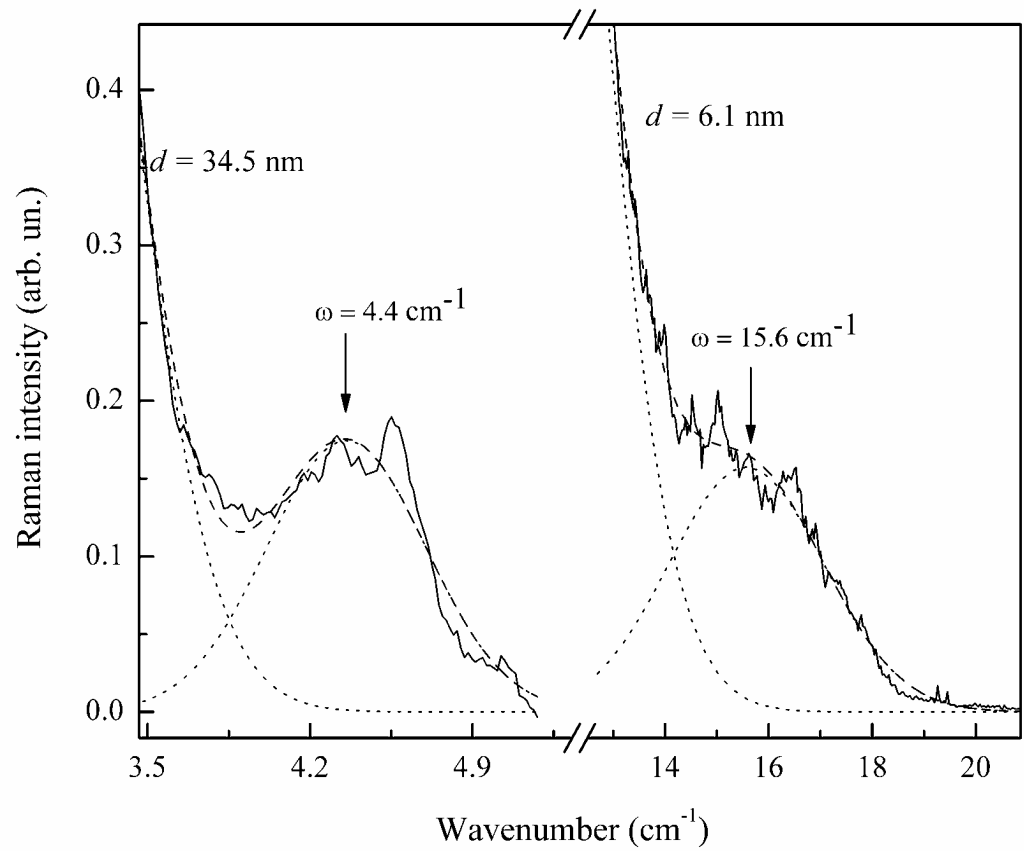

**Fig.6** Raman spectra of Cu nanoparticles in SiO$_2$ matrix measured at the temperature 293 K.

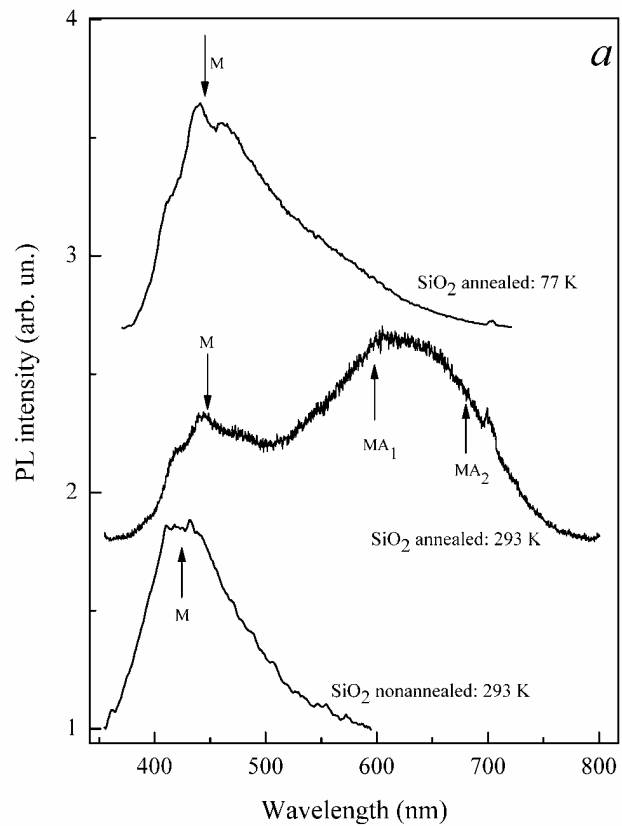 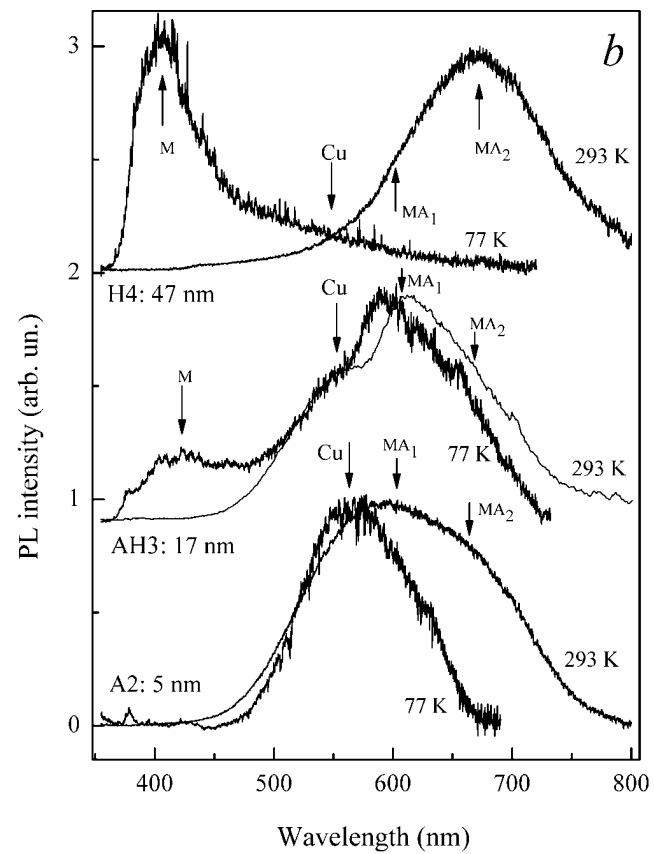

**Fig.7** (a) – Photoluminescence spectra of nonannealed and annealed undoped $SiO_2$ at the temperatures of 293 K and 77 K; index M marks the PL band of nonannealed $SiO_2$, $MA_{1,2}$ – the additional bands appearing in the annealed $SiO_2$. (b) – PL spectra of annealed $SiO_2$ matrix with the Cu nanoparticles of different sizes at the temperatures of 293 K and 77 K; index Cu marks the PL band of Cu nanoparticles.